\documentclass{iopjournal}

\usepackage{amssymb}
\usepackage{color}
\usepackage{bm}
\usepackage{threeparttable}
\usepackage{booktabs}
\usepackage{algorithmic}
\usepackage{cite}
\usepackage{textcomp}
\usepackage{stfloats}
\usepackage{url}
\usepackage{verbatim}
\usepackage{eqnarray}
\usepackage{graphicx}
\usepackage{balance}
\usepackage{graphics}
\usepackage[usenames,dvipsnames]{xcolor}
\usepackage{dcolumn}
\usepackage[utf8]{inputenc}
\usepackage{textcomp}
\usepackage{footnote}
\usepackage{tablefootnote}
\usepackage{hologo}
\usepackage{booktabs}
\usepackage{epstopdf}
\usepackage{url}
\usepackage{csquotes}
\usepackage[british]{babel}
\DeclareUnicodeCharacter{2212}{-}
\DeclareUnicodeCharacter{03B4}{$\delta$}
\usepackage{siunitx}
\usepackage[switch]{lineno}
\def\BibTeX{{\rm B\kern-.05em{\sc i\kern-.025em b}\kern-.08em
    T\kern-.1667em\lower.7ex\hbox{E}\kern-.125emX}}

\newcommand{\iint}{\mathop{\int\!\!\int}\nolimits}

\begin{document}

\articletype{Paper} 

\title{Massive Mitigation of Transport AC Losses in Superconducting Hybrid CORC–TSTC Cables}

\author{Hasan N. Al-Ssalih$^1$\orcid{0009-0001-6061-0965}, Antonio Badía-Majós$^2$\orcid{0000-0002-8753-2397} and Harold S. Ruiz$^{1}$\orcid{0000-0002-6100-1918}}

\affil{$^1$School of Engineering and Space Park Leicester, University of Leicester, LE1 7RH Leicester, U.K.}

\affil{$^2$~Instituto de Nanociencia y Materiales de Aragón (INMA), CSIC-Universidad de Zaragoza, C/ Pedro Cerbuna, 12 – 50009 Zaragoza, Spain.}



\begin{abstract}
High-current superconducting cables are emerging as key enablers for next-generation power transmission systems; however, their deployment is often limited by transport AC losses. Hybrid superconducting cables combining Conductor-on-Round-Core (CORC) and Twisted Stacked-Tape Conductor (TSTC) architectures have recently been proposed as a promising route toward power cables with high current capacity and compact form factors. However, their electrodynamic response under transport current operation remains poorly understood, particularly regarding how current injection conditions govern internal current redistribution. Here, we employ a three-dimensional electromagnetic model, previously validated against magnetisation experiments in equivalent cables, to investigate the influence of current injection strategy on the electrodynamic behavior of hybrid CORC-TSTC cables under self-field conditions. By comparing configurations in which the total current is either injected through a common connection between the CORC and TSTC conductors (non-insulated feeding) or supplied independently to each conductor (insulated feeding), we show that electrical coupling in non-insulated designs leads to strong current redistribution, inducing pronounced waveform distortion and elevated AC losses once the CORC layers approach magnetic saturation. In contrast, independent current feeding suppresses inter-conductor current exchange, stabilises the current waveforms, and exhibits an outstanding reduction in transport AC losses of up to 90\% at practical operating currents, compared with conventional feeding schemes. These findings reveal the central role of the current injection strategy in governing the internal electrodynamics and energy dissipation of hybrid superconducting cables, and identify the electrical decoupling of the constituent conductors at the feeding point as a simple and scalable route toward ultra-efficient hybrid superconducting power cables.
\end{abstract}



\section{Introduction}\label{Sec.1}
Superconducting technologies are increasingly recognised as critical enablers for the energy transition, advanced fusion devices, and next-generation power systems. In particular, second-generation high-temperature superconducting (2G-HTS) coated conductors provide extremely high current densities, enabling compact and efficient cables capable of delivering very large power levels within constrained footprints, a requirement that is becoming increasingly relevant for modern power delivery systems in data centres, power generation facilities, and utility networks. In this context, HTS cables can outperform conventional copper or aluminum solutions in both transmission and magnet applications~\cite{Coombs2024NatRev,FetisovProgSupAndCry2022,Driscoll_21,Mitchell_2021,ReyElsevier2015}. However, one of the major challenges preventing their widespread deployment lies in their AC loss performance, as such losses directly affect the cryogenic load, operational cost, and system stability~\cite{Sumption2022ac}. Therefore, understanding, predicting, and ultimately reducing AC losses is central to the design of practical HTS cables. However, AC losses are inherently linked to the underlying mechanisms of energy dissipation in superconducting tapes, which remain non-trivial to model accurately and are even more challenging to control or mitigate.

In practical devices, AC losses originate from the hysteretic behaviour of type-II superconductors when subjected to time-varying transport currents or magnetic fields. Therefore, their mitigation has become central to the design of advanced superconducting cable architectures. Within the strong pinning regime, this requires understanding the spatio-temporal dynamics of the current density inside the superconducting material~\cite{Ruiz2025-PMS}, as well as the mechanisms of current sharing via injection and induction between the individual conducting threads forming a cable. To address these challenges, several conductor architectures have been developed that balance transport capacity, mechanical robustness, and loss reduction. Among the most prominent approaches for large-scale deployment are the Conductor on Round Core (CORC) cable~\cite{Weiss2016SUST,VanDerLaan2019SUST} and the Twisted Stacked-Tape Conductor (TSTC)~\cite{TakayasuIOP2012,TakayasuIEEE2017}.

CORC cables, based on helically wound layers of REBCO tapes around a cylindrical former, offer excellent mechanical flexibility and robustness, while their transposed geometry helps mitigate coupling currents. In contrast, TSTCs adopt a twisted geometry of stacked tapes, leading to compact cross-sections, high current density, and relatively straightforward fabrication. Nevertheless, each design exhibits intrinsic trade-offs: CORC cables tend to be spatially inefficient owing to their required larger cross-section, whereas TSTC conductors, although compact, are limited in their total current-carrying capacity. Using full three-dimensional (3D) electromagnetic models, our previous work has shown that both configurations sustain very distinctive current density distributions that cannot be captured by simplified 2D or 3D-gauged approaches, while still reproducing experimentally consistent AC loss behaviour under realistic operating conditions~\cite{Ruiz2022-IEEETAS_CORC-3DFEM,Ruiz2024-JAP,Ruiz2025-IEEEAccess-TSTC}.

To combine the complementary advantages of both designs, a hybrid CORC–TSTC cable was recently proposed and experimentally demonstrated by Yoon and co-workers~\cite{YoonMiyeon2024ESoM}. In this architecture, an inner TSTC core is embedded within an outer CORC conductor, enabling a favourable balance among compactness, high current capacity, and mechanical robustness. Under transverse applied magnetic fields, the AC losses of such hybrid cables have recently been benchmarked against our fully three-dimensional electromagnetic models, which resolve the distribution of superconducting currents across micrometre-thick layers over cable lengths many orders of magnitude larger. Despite the substantial computational complexity involved compared with reduced or simplified modelling approaches, this level of fidelity is required to ensure full consistency with the governing laws of electromagnetism in type-II superconductors, and to provide a physically sound description of the macroscopic mechanisms governing the spatio-temporal dynamics of the critical current density within the superconducting tapes~\cite{Ruiz2025-PMS}. Using this approach, we reproduced the experimental AC loss measurements with excellent accuracy and revealed markedly different magnetisation current behaviours in the interacting CORC and TSTC architectures~\cite{Ruiz2025-IEEEAccess-Hybrid}.

In contrast, the electrodynamic response of hybrid CORC–TSTC cables under self-field transport current conditions remains largely unexplored experimentally, as calorimetric measurements of AC losses in sufficiently long prototypes are technically demanding, costly, and available only at a few facilities worldwide. In this context, validated three-dimensional models provide a unique opportunity to predict transport-current-driven losses and to extend the understanding of hybrid cable performance beyond the current experimental limitations. A central aspect of this problem is the injection of transport current into the hybrid conductor. Owing to the fundamentally different electromagnetic properties of CORC and TSTC architectures, the choice of feeding scheme directly determines how current is redistributed between them and how mutual inductive coupling develops. In the non-insulated configuration, both conductors share a common current source (\autoref{Fig_1}a), allowing inter-conductor current redistribution.
In the insulated configuration, each architecture is fed independently (\autoref{Fig_1}b), suppressing the current exchange between the CORC and TSTC parts.
As shown in this study, these two feeding schemes lead to strikingly different electrodynamic responses and AC loss characteristics of the hybrid cable.

The present work addresses these questions by developing a detailed three-dimensional finite-element model to analyse the transport current injection and redistribution in hybrid CORC–TSTC cables. By comparing insulated and non-insulated feeding configurations, we quantify the impact of current injection schemes on dynamic current sharing and AC loss behaviour. Crucially, this study demonstrates that insulated configurations can reduce the overall AC losses of the hybrid cable by up to nearly 90\% relative to more conventional designs in which the CORC and TSTC conductors are electrically connected, with the most pronounced benefit obtained when both conductors contain the same number of HTS tapes. These results provide new physical insight into the electrodynamics of hybrid superconducting cables and establish clear design guidelines for their deployment in high-power transmission and magnet applications.

The remainder of this paper is organised as follows. Section~\ref{Sec.2} introduces the numerical modelling framework and physical principles that outline the treatment of transport current injection under self-field conditions, including the definition of insulated and non-insulated feeding schemes. Section~\ref{Sec.3} presents and discusses the resulting electrodynamic behaviour of hybrid CORC–TSTC cables, examining how different current injection strategies govern transport current redistribution, waveform distortion, and the resulting AC losses. Finally, Section~\ref{Sec.4} summarises the main findings and highlights their implications for the design and optimisation of ultra-low-loss hybrid superconducting cables.

\section{Method}\label{Sec.2}

The dimensions of the hybrid CORC–TSTC cables analysed in this study strictly follow those of the experimental prototypes reported in Ref.~\citeonline{YoonMiyeon2024ESoM}, by combining two complementary conductor designs within a single architecture. The hybrid consists of an outer CORC conductor formed by six SuNAM REBCO tapes helically wound around a hollow cylindrical former, enclosing an inner TSTC core assembled from two to six twisted stacked tapes. Each SuNAM tape is $4$~\unit{mm} wide, with a $70$~$\mu$\unit{m} thick Hastelloy substrate and a self-field critical current of $190$~\unit{A} at 77~\unit{K}, resulting in nominal transport current capacities for the hybrid designs ranging from $1520$~\unit{A} to $2280$~\unit{A}.

The electromagnetic behaviour is computed using the three-dimensional partial differential equation module implemented in COMSOL Multiphysics, with the magnetic field strength vector $\mathbf{H}$ chosen as the primary state variable. This approach avoids any gauge-based reduction in the degrees of freedom in the magnetic flux density $\mathbf{B}$ across the superconducting tape dimensions. The same framework has been previously applied to CORC cables~\cite{Ruiz2024-JAP}, TSTCs~\cite{Ruiz2025-IEEEAccess-TSTC}, and hybrid conductors under transverse magnetisation fields~\cite{Ruiz2025-IEEEAccess-Hybrid}, where the full mathematical formulation of the governing PDE system, the non-linear $E$–$J$ constitutive law for the superconductor, and the magneto-angular dependent critical current density function $J_{c}(B,\theta)$ can be found, together with their excellent agreement with the experimental magnetisation loss measurements. The same set of equations and characteristic parameter values for the superconducting tapes is employed here. Accordingly, rather than revisiting the already benchmarked theoretical formulation, we focus on the key methodological distinction of the present study: the treatment of transport-current injection under self-field conditions, where the current-feeding scheme itself becomes the source of both transport and inductive losses.

\begin{figure}[!t]
\centering\resizebox{0.8\textwidth}{!}{\includegraphics{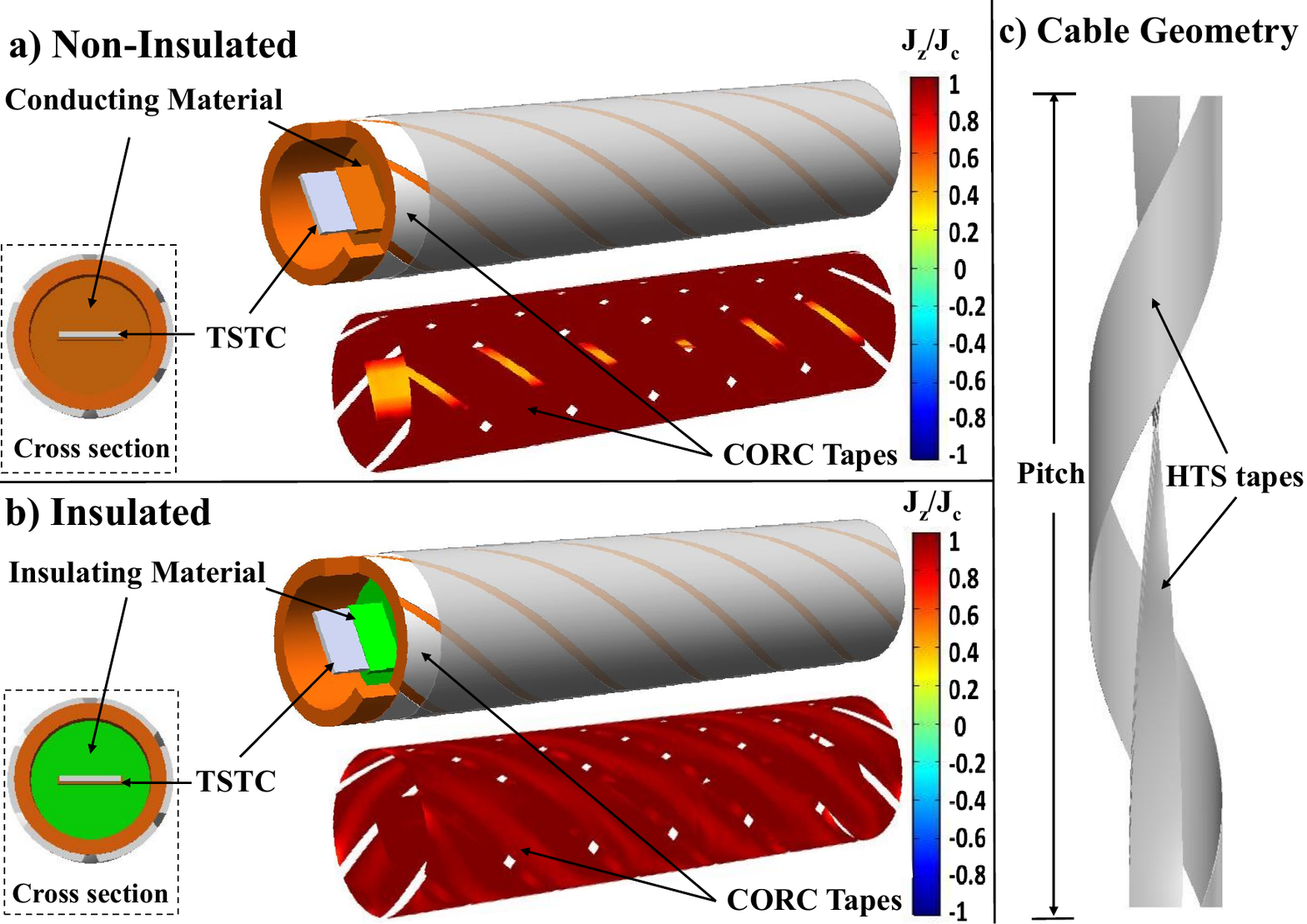}}
\caption{\label{Fig_1} {Pictorial representation of hybrid CORC-TSTC cables with current feeders in the so-called: \textbf{a)} \textit{non-insulated} configuration where CORC \& TSTC conductors are electrically connected in the terminals};  \textbf{b)} the \textit{insulated} configuration where the CORC and TSTC are electrically decoupled and; \textbf{c)} Simplified view of the helical winding characteristic of CORC Tapes, enclosing the twisted tapes of the TSTC along a pitch length.}
\end{figure}

Two transport current feeding schemes are considered, as illustrated in~\autoref{Fig_1}, differing in how the boundary condition for a full injected current of amplitude $I_{a}$ and angular frequency $\omega$ is imposed. With reference to the total current density $\mathbf{J}$ crossing a conductor of cross-sectional area $S$ along the outward normal direction $\mathbf{\hat{n}}$, the transport current is defined as
\begin{eqnarray}\label{Eq_1}
{I_{t} = I_{a}\,\sin{(\omega t)} = \iint \mathbf{J}\cdot \mathbf{\hat{n}}\, dS \,} .
\end{eqnarray}

On the one hand, in the \textit{non-insulated} configuration, the total transport current $I_{t}$ is applied as a single boundary constraint to the entire cable, allowing current redistribution between the CORC and TSTC architectures. In this case, all HTS tapes in the CORC and TSTC conductors are electrically connected and share a common current source through a single definition of $S$ in~\autoref{Eq_1}, enabling inter-conductor current exchange. However, in the \textit{insulated} configuration, independent boundary constraints are applied to the CORC and TSTC conductors, forcing the current to remain confined within each architecture and preventing any inter-conductor redistribution. In both cases, the total transport current is identical; however, the resulting current density distributions and therefore the associated AC losses are markedly different, as demonstrated in~\autoref{Sec.3}.

This distinction is critical because the CORC and TSTC architectures respond fundamentally differently to inductive currents. CORC cables tend to sustain helical magnetisation currents along their wound layers~\cite{Ruiz2024-JAP}, whereas TSTCs behave like twisted thin films, with current closure occurring primarily across their slab-like cross-section~\cite{Ruiz2025-IEEEAccess-TSTC}. As previously demonstrated under transverse magnetisation fields~\cite{Ruiz2025-IEEEAccess-Hybrid}, these contrasting behaviours are already apparent when the two conductors are analysed independently. When both architectures coexist within a single hybrid cable, the manner in which current is injected determines whether these effects reinforce or compete, ultimately shaping the overall AC loss response of the hybrid system.

\section{Results and Discussion}\label{Sec.3}

With a particular focus on the role of the current injection scheme in shaping the electrodynamic response of hybrid CORC–TSTC cables, we first examine how insulated and non-insulated feeding configurations govern the redistribution of transport current between the CORC and TSTC conductors, giving rise to markedly different current waveforms and distortion phenomena. Building on this analysis, we then quantify how these redistribution mechanisms translate into transport AC losses, demonstrating that electrical insulation at the feeding point can dramatically suppress energy dissipation at high operating currents.

\subsection{Effect of Current Injection Scheme on Transport Current Distribution}\label{Sec.3.1}

We begin by analysing the transport current distribution within the HTS tapes of the hybrid cable configurations shown in~\autoref{Fig_1}. Two feeding schemes are considered: (i) a \emph{non-insulated} configuration, in which the CORC and TSTC conductors are electrically connected to each other at the current injection point and supplied by a common current source, and (ii) an \emph{insulated} configuration, in which the two conductors are electrically decoupled, including at the current injection point.

\begin{figure}[!t]
\centering
\resizebox{0.8\textwidth}{!}{\includegraphics{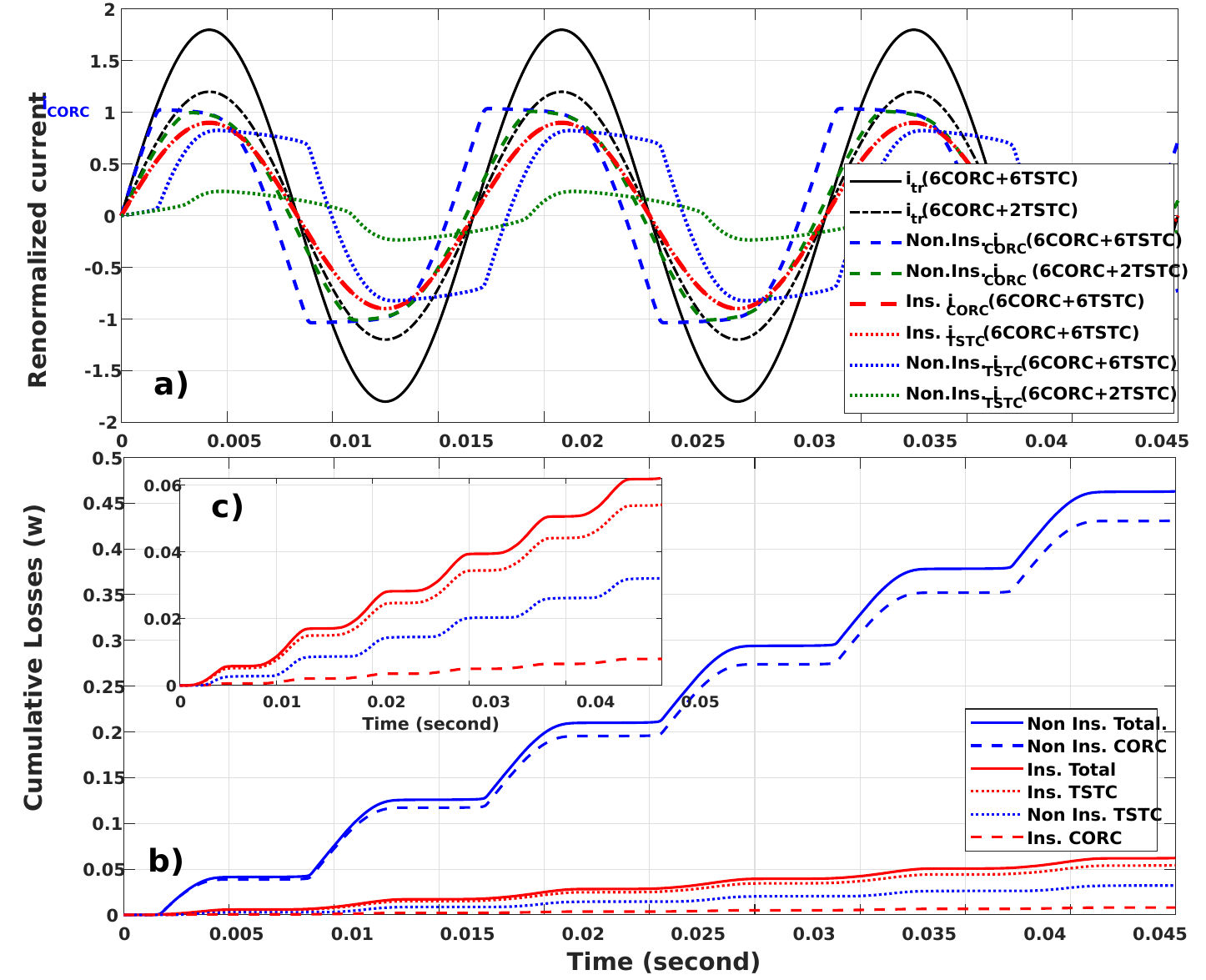}}
\caption{\label{Fig_2}
{Simulated AC loss and transport current distribution in 6-by-6 and 6-by-2 hybrid CORC-TSTC cables subjected to an applied transport current equal to $90\%$ of the cable’s rated capacity, i.e., $I_{a}=0.9\times n_\text{tapes}\times I_{c}$, at a frequency of 60~{Hz}, and ${I_{c}=190}$~\unit{A} at 77~\unit{K}.
(a) Transport current response for the non-insulated and insulated configurations normalized by the theoretical maximum current allowed at the CORC conductor ($I_{CORC}=1140$~\unit{A}) ;
(b) Exemplary cumulative losses for the 6-by-6 hybrid cables in the insulated and non-insulated configurations, highlighting the individual loss contributions from each conductor. (c) Shows a zoom in of the relevant losses displayed at (b).}
}
\end{figure}

In the \textbf{non-insulated configuration}, both conductors are electrically coupled and must collectively satisfy the single transport-current constraint in~\autoref{Eq_1}. This ensures that the total applied current does not exceed the overall cable capacity, that is, $I_{a} \leq n_{\text{tapes}}\langle I_{c}\rangle$, where $n_{\text{tapes}}$ denotes the total number of HTS tapes in the hybrid cable and $\langle I_{c}\rangle$ their average self-field critical current. For comparison, the transport current in~\autoref{Fig_2}(a) is normalised by the theoretical maximum current of the CORC conductor, $I_{\mathrm{CORC}} = 1140$~\unit{A}, such that $i_{t}=I_{t}/I_{\mathrm{CORC}}$. Under non-insulated feeding, the transport current distributes non-uniformly between the two conductors. The outer CORC conductor (blue and green dashed lines for hybrid cables with 6- and 2-tape TSTC cores, respectively) carries a larger fraction of the current than the inner TSTC conductor (blue and green dotted lines). This behaviour arises from current sharing driven by the cylindrical geometry of the hybrid cable, where the energetic cost associated with magnetisation currents favours current flow in the outer regions of the superconducting structure~\cite{Ruiz2025-PMS}. Consequently, the current preferentially occupies the CORC layers, limiting the current flowing in the TSTC core.

Once the CORC tapes approach saturation, any additional transport current is diverted progressively to the TSTC conductor, as evidenced in~\autoref{Fig_2}. This redistribution is analogous to the classical interplay between transport and magnetisation currents in cylindrical superconducting wires~\cite{Ruiz2012-APL}. When the current approaches the critical threshold of the CORC conductor, pronounced semi-plateaus emerge in the time evolution of the TSTC current, reflecting the constrained ability of the CORC layers to accommodate further current. Under these conditions, waveform distortion and phase shifts become apparent in the TSTC conductor, particularly when the applied current exceeds approximately 50\% of the rated capacity of the cable. These effects are most pronounced for hybrid cables containing a larger number of TSTC tapes.
At approximately 90\% of $i_{t}$, when the CORC layers operate close to their critical current density and no flux-free region remains, the distortion of the TSTC current waveform becomes particularly severe (see dotted curves in~\autoref{Fig_2}(a)).

In contrast, when the \textbf{CORC and TSTC conductors are electrically insulated}, the transport current us distributed uniformly between them, as indicated by the dash-dotted curves in~\autoref{Fig_2}(a). In this case, the constraint in~\autoref{Eq_1} is applied independently to each conductor, forcing the injected current to remain confined within each architecture. Therefore, the total transport current is the sum of the independently supplied currents to the CORC and TSTC conductors, both of which may be interpreted as parallel single-phase conductors driven in phase and connected in parallel at the terminals. However, it is worth mentioning that in practice, this configuration would require compatible voltage compliance among the current sources to ensure stable current sharing and to avoid the overloading of individual conductors. Under this condition, the insulated-conductor configuration suppresses inter-conductor current redistribution and preserves balanced current waveforms without significant distortion or phase shift. Yet, while non-insulated feeding may introduce waveform distortion and reduced signal stability at conductor terminals, its impact on the overall energy dissipation remains to be quantified. This aspect is addressed explicitly through AC loss analysis in~\autoref{Sec.3.2}.
\subsection{Transport AC Loss Reduction in Insulated Hybrid Cables}\label{Sec.3.2}

\begin{figure}[!t]
\centering\resizebox{0.8\textwidth}{!}{\includegraphics{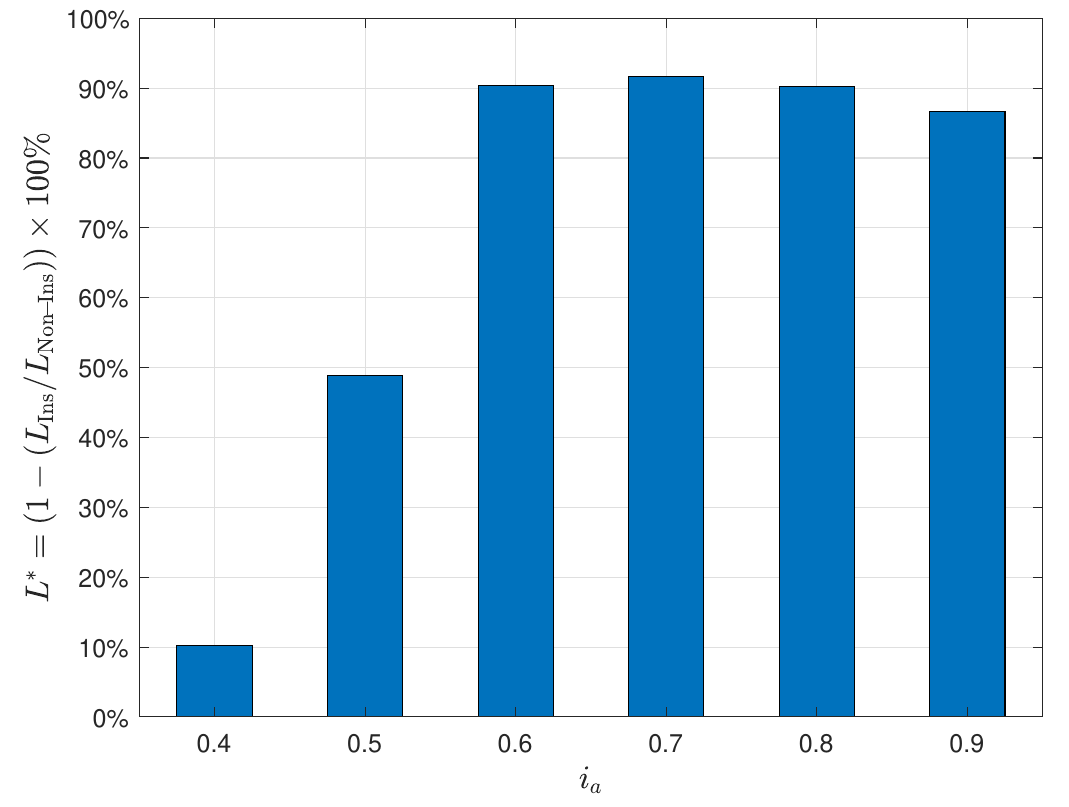}}
\caption{\label{Fig_3} {Percentual loss reduction of the $6$ by $6$ insulated CORC-TSTC hybrid cable under transport current with respect to a non-insulated hybrid cable with the same number of tapes.}}
\end{figure}

The presence of electrical insulation between the CORC and TSTC conductors plays a decisive role in determining the AC transport losses of hybrid cables, particularly at high transport currents. ~\autoref{Fig_3} summarises the relative loss reduction achieved by insulated configurations, defined as $L^{*}=1-(L_{\mathrm{Ins}}/L_{\mathrm{Non\text{-}Ins}})$ in percentage units. Insulated hybrid cables exhibit an AC loss reduction of approximately 50\% once the applied transport current exceeds approximately half of the cable rated capacity ($ i_{a}\gtrsim 0.5$), with the reduction approaching nearly 90\% for currents above $0.7\,i_{a}$. This latter regime corresponds to the minimum practical operating range typically targeted in superconducting power applications and marks the onset of strong coupling between the current redistribution and energy dissipation mechanisms.

The origin of this behaviour becomes clear when analysing the cumulative AC loss contributions as shown in~\autoref{Fig_2}(b,c). In non-insulated hybrid cables, the absence of electrical isolation allows dynamic current redistribution between the CORC and TSTC conductors once the CORC tapes approach full magnetic flux penetration. When no flux-free region remains within the CORC layers, additional transport current can no longer be accommodated without a significant energetic penalty, leading to a progressive diversion of the current into the TSTC conductor. This redistribution is accompanied by a pronounced waveform distortion in the individual conductor currents, even though the combined current delivered by the source remains sinusoidal (see~\autoref{Fig_2}(a)). Such behaviour is characteristic of the interplay between transport and magnetisation currents in cylindrical superconducting geometries~\cite{Ruiz2012-APL}.

In contrast, insulating the CORC and TSTC conductors suppresses inter-conductor current exchange and enforces independent current evolution in each architecture. In this configuration, the transport current in each conductor closely follows the waveform of the power source, preventing the onset of distortion and phase shifts observed in the non-insulated case. Moreover, insulation enables a more effective utilisation of the transport current capacity of the CORC conductor, as magnetisation currents are dynamically accommodated within the same conductor rather than being transferred to the TSTC core. This mechanism is responsible for the substantial reduction in the total AC losses observed in insulated hybrid cables.

\begin{figure}[!t]
\centering
\resizebox{0.8\textwidth}{!}{\includegraphics{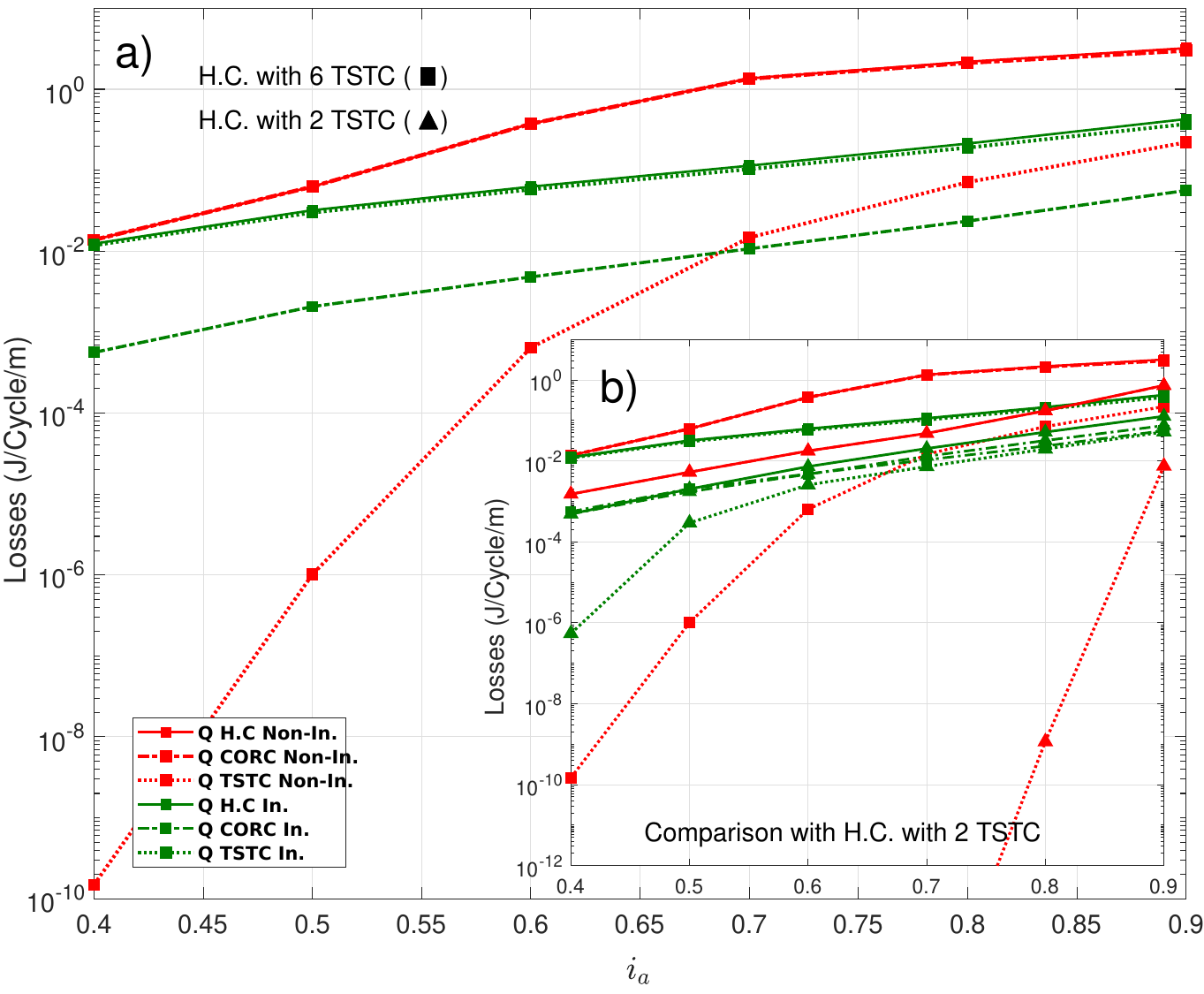}}
\caption{\label{Fig_4}
{a) AC losses as function of the normalized rated current of the hybrid CORC-TSTC cable, $i_{a}=I_{a}/(n_{tapes}\times<I_{c}>)$, for non-insulated (red curves) and insulated (green curves) layouts (\autoref{Fig_1}), each comprising six 2G-HTS tapes in the CORC conductor and within; a TSTC conductor made of up to (a) six 2G-HTS tapes -squares-, that are displayed for comparison against (b) a compelling hybrid cable with two 2G-HTS tapes -triangles-.}
}
\end{figure}

These effects become particularly evident at high transport currents. For non-insulated configurations, once the transport current exceeds the CORC critical threshold ($I_{\mathrm{CORC}}=1140$~\unit{A}), localised increases in AC losses emerge, as highlighted by the red dashed curves in~\autoref{Fig_4}(a). In this regime, the CORC conductor dominates the total energy dissipation for both hybrid designs considered, irrespective of whether the TSTC core contains two or six tapes. When the losses are examined separately, the TSTC conductor in the non-insulated hybrids consistently exhibits lower AC losses owing to its reduced current share (see~\autoref{Fig_4}(b), red dotted curves). However, this apparent advantage is offset at the cable level by the sharp increase in CORC losses once current redistribution becomes significant, leading to pronounced loss growth at approximately $0.6\,I_{t}$ and $0.9\,I_{t}$ for the two- and six-tape TSTC configurations, respectively.

In contrast, insulated hybrid cables maintain nearly identical CORC loss levels across configurations (see~\autoref{Fig_4}(b), green dashed curves), reflecting the improved current homogeneity enforced by insulation. Overall, these results demonstrate that electrically insulating the CORC and TSTC conductors at the feeding point provides an exceptionally effective and practical route to suppress transport AC losses in hybrid cables, outperforming alternative strategies for cables with comparable current ratings.

\section{Conclusions}\label{Sec.4}

In this article, we present a comprehensive three-dimensional electromagnetic modelling study that addresses the critical role of conductor insulation in hybrid CORC–TSTC superconducting cables operating under self-field transport current conditions. By explicitly distinguishing between insulated and non-insulated current injection schemes, we analyse their impact on the transport current distribution, waveform distortion, and striking AC loss behaviour under standard AC operating conditions.

For non-insulated configurations, once the applied transport current exceeds the nominal CORC current ($I_{\mathrm{CORC}}$), pronounced distortions in both waveform shape and phase emerge in the currents flowing through the TSTC conductor. These effects intensify as the transport current increases beyond approximately 70\% of the cable rated capacity, reflecting the onset of strong current redistribution driven by magnetic flux penetration in the CORC layers. In contrast, insulated configurations enforce independent current evolution in each conductor, resulting in nearly identical and undistorted current waveforms with no inter-conductor current sharing.

Beyond current distribution, conductor insulation has a decisive impact on transport AC losses. In insulated hybrid cables, although the TSTC conductor contributes a larger fraction of the total losses, the overall AC loss of the cable is substantially reduced without altering the loss profile of the CORC conductor. Notably, the CORC losses remain nearly unchanged irrespective of the number of HTS tapes in the TSTC core, highlighting the robustness of the outer conductor under insulated feeding conditions.

Among all configurations examined, insulated hybrid cables consistently exhibit the lowest transport AC losses, whereas the non-insulated configurations display the highest losses once the applied current exceeds $I_{\mathrm{CORC}}$, with losses increasing rapidly with current amplitude. While the TSTC conductor in non-insulated designs exhibits comparatively low losses owing to its reduced current share, this behaviour ultimately reflects an inefficient utilisation of the hybrid architecture when conductor insulation is not considered.

Overall, these results demonstrate that a physically grounded, three-dimensional description of the current density dynamics is essential for understanding and optimising the electrodynamic performance of hybrid superconducting cables. For hybrid CORC–TSTC designs in particular, electrically insulating the constituent conductors at the feeding point emerges as a simple yet highly effective strategy to achieve exceptional performance under practical operating currents. Importantly, this approach can be readily implemented through an appropriate current feeder design, offering a clear and scalable pathway for the deployment of low-loss hybrid superconducting cables in high-power applications.
\ack{}
This work was supported in part by the U.K. Research and Innovation (UKRI), Engineering and Physical Sciences Research Council (EPSRC), under grant Ref. EP/S025707/1 led by H.S.R. The work of H.N.H.A. was supported by Iraq’s Higher Committee for Education Development (HCED). The work of A. B-M. was supported by the projects PID2023-146041OB-C21 (funded by {MICIU/AEI/10.13039/501100011033} and ERDF/EU) and T54-23R (funded by Gobierno de Aragón).

\roles{Author contributions statement}
Conceptualization: H.N.A, H.S.R;
Methodology: H.N.A, H.S.R.; Investigation: H.N.A, H.S.R.; Visualization: H.N.A., H.S.R., A.B.M.; Results Analysis: H.S.R., A.B.M.; Funding acquisition: H.S.R., A.B.M.; Project administration: H.S.R.; Supervision: H.S.R.; Writing – original draft: H.N.A.; Writing – review \& editing: H.S.R., A.B.M.

\section*{Additional information}

\textbf{Competing interests:} The authors have no competing interests to declare.

\noindent\textbf{Data and materials availability:} All the data needed to evaluate and reproduce the Results and Discussion in the article are accessible through its contents and previous work in Ref.\cite{Ruiz2025-IEEEAccess-Hybrid}.

\noindent\textbf{License information:} For the purpose of open access, the authors have applied a Creative Commons Attribution (CC BY) licence to the Author Accepted Manuscript version arising from this submission.

\bibliographystyle{iopart-num}
\bibliography{Main_References}

@article{VanDerLaan2019SUST,
issn = {0953-2048},
journal = {Superconductor Science and Technology},
pages = {33001},
volume = {32},
publisher = {IOP Publishing},
number = {3},
year = {2019},
title = {{Status of CORC cables and wires for use in high-field magnets and power systems a decade after their introduction}},
url = {https://doi.org/10.1088/1361-6668/aafc82},
copyright = {2019 IOP Publishing Ltd},

address = {United States},
author = {van der Laan, D C and Weiss, J D and McRae, D M},
doi = {https://doi.org/10.1088/1361-6668/aafc82},
organization = {Univ. of Colorado, Boulder, CO (United States)},
}

@Article{Coombs2024NatRev,
author={Coombs, Tim A.
and Wang, Qi
and Shah, Adil
and Hu, Jintao
and Hao, Luning
and Patel, Ismail
and Wei, Haigening
and Wu, Yuyang
and Coombs, Thomas
and Wang, Wei},
title={High-temperature superconductors and their large-scale applications},
journal={Nature Reviews Electrical Engineering},
year={2024},
month={Dec},
day={01},
volume={1},
number={12},
pages={788-801},
issn={2948-1201},
doi={10.1038/s44287-024-00112-y},
 url={https://doi.org/10.1038/s44287-024-00112-y}
}

@incollection{Sumption2022ac,
  title={{AC} losses in superconducting materials, wires, and tapes},
  author={Sumption, Michael D and Majoros, Milan and Collings, Edward W},
  booktitle={Handbook of Superconductivity},
  pages={238--250},
  year={2022},
  publisher={CRC Press}
}

@article{YoonMiyeon2024ESoM,
issn = {1051-8223},
journal = {IEEE Transactions on Applied Superconductivity},
pages = {1--5},
volume = {34},
publisher = {IEEE},
number = {5},
year = {2024},
title = {{Experimental Study of Magnetization Loss by External Magnetic Field in CORC-TSTC Hybrid Composite Conductor}},
url = {https://doi.org/10.1109/TASC.2023.3347376},
copyright = {Copyright 2024 Elsevier B.V., All rights reserved.},

address = {New York},
author = {Yoon, Miyeon and Lee, Myeonghee and Lee, Ji-Kwang and Choi, Kyeongdal and Kim, Woo-Seok},
}

@incollection{ReyElsevier2015,
pages = {138--145},
publisher = {Elsevier Ltd},
booktitle = {{Superconductors in the Power Grid: Materials and Applications}},
isbn = {9781782420378},
year = {2015},
title = {{High-temperature superconducting ({HTS}) {AC} cables for power grid applications}},
uri={https://doi.org/10.1016/B978-1-78242-029-3.00005-4},
copyright = {2015 Elsevier Ltd},

author = {Malozemoff, A.P. and Yuan, J. and Rey, C.M.},
series = {Woodhead Publishing series in energy}
}

@article{FetisovProgSupAndCry2022,
author  = {Fetisov, S. S. and Zubko, V. V. and Nosov, A. A. and Zanegin, S. Yu. and Vysotsky, V. S.},
doi={https://doi.org/10.9714/psac.2020.22.4.031},
title = {{Review of the design, production and tests of compact AC HTS power cables}},
url={https://doi.org/10.9714/psac.2020.22.4.031},
journal = {Progress in Superconductivity and Cryogenics},
publisher = {한국초전도저온공학회},
volume = {22},
number = {4},
pages = {31-39},
year = {2020},
}

@article{TakayasuIEEE2017,
issn = {1051-8223},
journal = {IEEE Transactions on Applied Superconductivity},
pages = {1--5},
volume = {27},
publisher = {IEEE},
number = {4},
year = {2017},
title = {{Investigation of HTS Twisted Stacked-Tape Cable (TSTC) Conductor for High-Field, High-Current Fusion Magnets}},
url = { https://ieeexplore-ieee-org.ezproxy3.lib.le.ac.uk/stamp/stamp.jsp?tp=&arnumber=7815291},

address = {United States},
author = {Takayasu, Makoto and Chiesa, Luisa and Noyes, Patrick D. and Minervini, Joseph V.},
doi = {https://doi.org/10.1109/TASC.2017.2652328},
organization = {Massachusetts Inst. of Technology (MIT), Cambridge, MA (United States)},
}

@article{TakayasuIOP2012,
issn = {0953-2048},
journal = {Superconductor Science and Technology},
pages = {14011--1-21},
volume = {25},
number = {1},
year = {2012},
title = {{HTS twisted stacked-tape cable conductor}},
url = {http://stacks.iop.org/SUST/25/014011},

author = {Takayasu, Makoto and Chiesa, Luisa and Bromberg, Leslie and Minervini, Joseph V},
doi = {https://doi.org/10.1088/0953-2048/25/1/014011},
}

@article{Driscoll_21,
author = {MacManus-Driscoll, Judith L. and Wimbush, Stuart C.},
journal = {Nature Reviews Materials},
number = {7},
pages = {587--604},
title = {Processing and application of high-temperature superconducting coated conductors},
volume = {6},
url = {https://doi.org/10.1038/s41578-021-00290-3},
year = {2021}
}

@article{Mitchell_2021,
url = {https://doi.org/10.1088/1361-6668/ac0992},
year = {2021},
month = {sep},
publisher = {IOP Publishing},
volume = {34},
number = {10},
pages = {103001},
author = {Neil Mitchell and Jinxing Zheng and Christian Vorpahl and Valentina Corato and Charlie Sanabria and Michael Segal and Brandon Sorbom and Robert Slade and Greg Brittles and Rod Bateman and Yasuyuki Miyoshi and Nobuya Banno and Kazuyoshi Saito and Anna Kario and Herman Ten Kate and Pierluigi Bruzzone and Rainer Wesche and Thierry Schild and Nikolay Bykovskiy and Alexey Dudarev and Matthias Mentink and Franco Julio Mangiarotti and Kamil Sedlak and David Evans and Danko C Van Der Laan and Jeremy D Weiss and Min Liao and Gen Liu},
title = {Superconductors for fusion: a roadmap},
journal = {Superconductor Science and Technology}
}

@article{Weiss2016SUST,
issn = {0953-2048},
journal = {Superconductor Science and Technology},
pages = {14002},
volume = {30},
publisher = {IOP Publishing},
number = {1},
year = {2016},
title = {Introduction of CORC wires: highly flexible, round high-temperature superconducting wires for magnet and power transmission applications},
url= {https://doi.org/10.1088/0953-2048/30/1/014002},
copyright = {2016 IOP Publishing Ltd},

address = {United Kingdom},
author = {Weiss, Jeremy D and Mulder, Tim and ten Kate, Herman J and van der Laan, Danko C}
}

@article{Ruiz2012-APL,
author = {H. S. Ruiz and A. Bad\'{\i}a-Maj\'{o}s and Y. A. Genenko and H. Rauh and S. V. Yampolskii},
title = {Superconducting wire subject to synchronous oscillating excitations: Power dissipation, magnetic response, and low-pass filtering},
journal = {Applied Physics Letters},
volume = {100},
number = {11},
pages = {112602},
year = {2012},
url = {https://doi.org/10.1063/1.3693614}
}

@article{Ruiz2022-IEEETAS_CORC-3DFEM,
issn = {1051-8223},
journal = {IEEE Transactions on Applied Superconductivity},
pages = {1--5},
volume = {32},
publisher = {IEEE},
number = {4},
year = {2022},
title = {{3D FEM Modeling of Commercial Cables With Bean's Like Magnetization Currents and Its AC-Losses Behavior}},
url={https://doi.org/10.1109/TASC.2022.3145309},

address = {New York},
author = {Fareed, M. U. and Kapolka, M. and Robert, B. C. and Clegg, M. and Ruiz, H. S.},
doi={ https://doi.org/10.1109/TASC.2022.3145309}
}

@article{Ruiz2024-JAP,
	author = {Clegg, M. and Ruiz, H. S.},
	doi = {https://doi.org/10.1063/5.0218241},
	issn = {0021-8979},
	journal = {Journal of Applied Physics},
	month = {07},
	number = {3},
	pages = {033902},
	title = {{3D electromagnetic assessment of bended CORC{\textregistered} cables}},
	url = {https://doi.org/10.1063/5.0218241},
	volume = {136},
	year = {2024}
}

@ARTICLE{Ruiz2025-IEEEAccess-TSTC,
  author={Clegg, Matthew and Al-Ssalih, Hasan N. H. and Ruiz, Harold S.},
  journal={IEEE Access}, 
  title={{Magnetization Losses of Transposed Stacked Tape Conductors Under AC Transverse Magnetic Fields}}, 
  year={2025},
  volume={13},
  number={},
  pages={84013-84023},
  url={http://doi.org/10.1109/ACCESS.2025.3566645},
}

@Article{Ruiz2025-PMS,
title = {{Critical Current Density in Advanced Superconductors}},
journal = {Progress in Materials Science},
pages = {101492},
year = {2025},
issn = {0079-6425},
doi = {https://doi.org/10.1016/j.pmatsci.2025.101492},
url = {https://doi.org/10.1016/j.pmatsci.2025.101492},
author = {Ruiz, H. S. and H\"{a}nisch, J. and Polichetti, M. and Galluzzic, A. and Gozzelino, L. and Torsello, D. and Milo\v{s}evi\'{c}-Govedarovi\'{c}, S. and Grbovi\'{c}-Novakovi\'{c}, J. and Dobrovolskiy, O.V. and Lang, W. and Grimaldi, G. and Crisan, A. and Badica, P. and Ionescu, A.M. and Cayado, P. and Willa, R. and Barbiellini, B. and Eley, S. and Bad\'{\i}a–Maj\'{o}s, A.}
}

@article{Ruiz2025-IEEEAccess-Hybrid,
pages = {186952-186964},
publisher = {IEEE},
title = {Magnetization Losses and {Non-Reduced 3D Modeling of Hybrid CORC-TSTC Composite Conductors}},
doi = {https://doi.org/10.1109/ACCESS.2025.3627080},
url = {https://doi.org/10.1109/ACCESS.2025.3627080},
volume = {13},
author = {Al-Ssalih, Hasan N. H. and Clegg, Matthew and Badia-Majos, Antonio and Ruiz, Harold S.},
address = {Piscataway},
copyright = {Copyright 2025 Elsevier B.V., All rights reserved.},
issn = {2169-3536},
journal = {IEEE access},

year = {2025},
}
\end{document}